\documentclass[prl,showpacs,showkeys, superscriptaddress,amsfonts,amssymb,amsmath, twocolumn]{revtex4}
\usepackage{graphicx}
\usepackage{graphics}
\usepackage{dsfont}
\usepackage{amsfonts}
\usepackage{amssymb}
\usepackage{amsmath}
\allowdisplaybreaks[4]
\newcommand{\Name}[1]{{#1},}
\newcommand{\REVIEW}[4]{{\it #1}, {\bf #2} {(#3)} {#4}}

\begin{document}

\title{Hidden $Sp(2s\!+\!1)$- or $SO(2s\!+\!1)$-symmetry
and new exactly solvable models\\ in ultracold atomic systems}
\author{Yuzhu Jiang}
\affiliation{Beijing National Laboratory for Condensed
Matter Physics, Institute of Physics, Chinese Academy of Sciences,
Beijing 100190, People's Republic of China}
\affiliation{Technical Institute of Physics and Chemistry, Chinese Academy of Sciences,
Beijing 100190, People's Republic of China}

\author{Junpeng Cao}
\affiliation{Beijing National Laboratory for Condensed
Matter Physics, Institute of Physics, Chinese Academy of Sciences,
Beijing 100190, People's Republic of China}

\author{Yupeng Wang}

\affiliation{Beijing National Laboratory for Condensed
Matter Physics, Institute of Physics, Chinese Academy of Sciences,
Beijing 100190, People's Republic of China}

\begin{abstract}
The high spin ultracold atom models with a special form
of contact interactions, i.e., the scattering lengthes in the total
spin-$2,4 \cdots$ channels are equal but may be different from that
in the spin-0 channel, is studied. It is found that those models 
have either $Sp(2s+1)$-symmetry for the fermions or 
$SO(2s+1)$-symmetry for the bosons in the spin sector. Based on the
symmetry analysis, a new class of exactly solvable models is
proposed and solved via the Bethe ansatz. The ground states for
repulsive fermions are also discussed.
\end{abstract}



\maketitle

\section{Introduction}

Recently, the study on cold atoms with high spin has aroused lots of
attention in the fields of atomic, molecular, optical and condensed
matter physics. Due to the spin exchange interactions, many
interesting spin ordered states arise and the phase diagrams of
these systems are very rich. For instance, in the spin-1 spinor
Bose--Einstein condensations, the bosons are found to form the pairs
and the pairs condense even in the repulsive regime
\cite{Law1998PRL, Mukerjee2006PRL, Mueller2006PRA}. In experiments,
by using the atom cooling and trapping techniques, one can prepare
the high spin cold atomic systems, such as $^7$Li \cite{Bradley1995PRL}, $^{23}$Na \cite{Stamper-Kurn1998PRL1}, $^{87}$Rb \cite{Myatt1997PRL,
Barrett2001PRL, Paredes2004N} with hyperfine spin 1; $^{53}$Cr
\cite{Chicireanu2006PRA} with hyperfine spin 3/2; and $^{40}$K
\cite{DeMarco1999S}, $^{173}$Yb \cite{Takasu2006LP}, $^{43}$Ca
\cite{Witte1992JOSAB}, $^{87}$Sr \cite{Xu2003JOSAB}, $^{133}$Cs
\cite{Soding1998PRL, Ma2004JPB} with more higher ones. Using the
Feshbach resonance \cite{Inouye1998N, Dickerscheid2005PRA} and
confinement induced resonance \cite{Bergeman2003PRL} technique, the
interactions among the atoms can be manipulated. This provides a
good platform for studying the tunable condensed matter systems. In
theoretical approaches, the low-energy effective models of the
dilute ultracold atomic systems are the quantum gas with contact
interactions, and the spin exchanging interactions should also be
considered for systems with internal degrees of freedom
\cite{Ho1998PRL, Ohmi1998JPSJ}.

Symmetric analysis plays a very important role in studying the
quantum many-body systems. Physical properties such as ground state
manifold and order parameters are closely related to the symmetry of
a system \cite{WuCJ2005PRL, ChenS2005PRB, WuCJ2006MPLB}. The
analysis of the symmetry can give us some hints to do the suitable
approximation and study the physics such as phase diagram in the
frame of mean-field theory. It can also simplify the analytical and
numerical calculations. In the cold atomic systems with delta
function interactions, the symmetric or anti-symmetric properties of
the identical particles restrict the forms of spin exchange
interactions. Effective spin exchanging interactions only take place
in the channels with symmetric spatial wave function. Such property
may make the systems to have intrinsic symmetries in the spin
sector. For example, in the spin-$3/2$ system, the $SO(5)$ symmetry
is found \cite{WuCJ2003PRL}.

The strong quantum fluctuation and correlation make the physics of
one-dimensional (1D) system quite different from the ones of higher
dimensions. Many numerical and analytical methods are developed to
study the 1D systems. The exact solution is a good starting point to
study these systems, since it can give us conclusive results. A
well-known exactly solvable system in the cold atomic system is the
Lieb--Liniger model \cite{LiebLiniger1963PR}, where the
scalar bosons are studied. The fermion case with
spin-1/2 are exactly solved by Yang \cite{Yang1967PRL}.
Sutherland generalized Yang's model to arbitrary spin case, where
the system has the $SU(2s+1)$ symmetry \cite{Sutherland1968PRL},
with $s$ the spin of the particles. At present, it is clear that the
multicomponent quantum gas including the Bose--Fermi mixtures with
delta function interactions are integrable, if the masses of each
species are equal and the interactions are equal \cite{Zhou1988JPA1,
Lai1971PRA, Pu1987JPA}. In
these models, the scattering lengthes in different channels are the
same, for that the spin exchange interactions are not considered.
However, the spin exchanging usually can not be neglected in
experiments, and many novel ordered states are induced by the spin
exchanging. Motivated by this consideration, we proposed a $SO(3)$
integrable spin-1 bosonic model \cite{Cao2007EPL} and a $Sp(4)$
integrable spin-3/2 fermionic model \cite{Jiang2009EPL}, where the
contact spin exchange interactions are considered. The scattering
lengthes in different channels are different.

This paper is a generalization of our previous works
\cite{Cao2007EPL,Jiang2009EPL} to cold atom systems with arbitrary
hyperfine spins. For a special interaction form of atoms with
hyperfine spin $s$, we find that the fermion system has the $U(1)
\otimes Sp(2s+1)$ symmetry while the boson system has the $U(1)
\otimes SO(2s+1)$ symmetry. The generators of corresponding algebra
are constructed by the magnetic multipole operators. Based on the
symmetry analysis, we propose a new class of exactly solvable models
in one dimension.

\section{Model and symmetry}

For the delta function interaction models of dilute cold atomic gas
with hyperfine spin $s$, the spin exchange interaction between two
particles $i$ and $j$ is usually written as spin projection
operators $\hat P^l_{ij}$ in different channels with total spin-$l$
($l=0,1,2,\cdots,2s$).  Nontrivial scattering processes occur only
in the even $l$ channels because of the symmetry or anti-symmetry
behavior of the wave functions. To study the behavior of such
systems away from the $SU(2s+1)$ symmetry point, we consider a
simple case, i.e., all the scattering lengths of nonzero $l$
channels are the same. The Hamiltonian reads
\begin{equation}
\label{H} \hat H=- \sum_{i=1}^N \nabla ^2_{{\boldsymbol r}_j} +
\sum_{i\neq j}^N \Big[c_1\hat P^0_{ij} + c_2 \!\!\sum_{l=2,4,
\cdots}\!\! \hat P^l_{ij}\Big] \delta({\boldsymbol r}_i-{\boldsymbol
r}_j).
\end{equation}
Here, $N$ is the number of atoms, ${\boldsymbol r}_i$ is the
position of the $i$-th atom, $c_1$ is the interaction strength in
spin-0 channel and $c_2$ is the one in other channels.

\begin{figure}[t]
\begin{center}
\includegraphics[width = \linewidth]{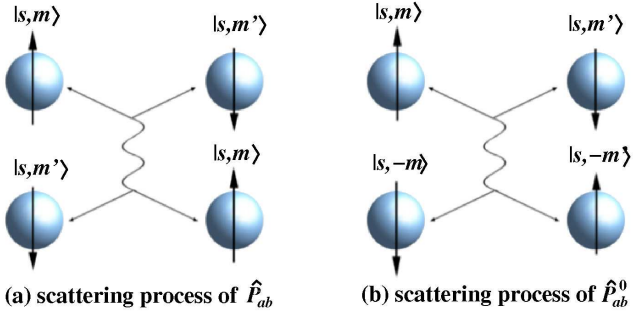}
\end{center}
\caption{The sketches of two-atom scattering processes.\label{fig1}}
\end{figure}

The two-body scattering in the system (\ref{H}) is quite
interesting. There are two kinds of scattering processes in the spin
sector. One is $|s,m;s,m'\rangle \langle s,m';s,m| + h.c.$ provided
by two-particle permutation as shown in Fig. \ref{fig1}(a), here $m$
and $m'$ are the spins along $z$ direction, and $m,
m'=s,s-1,\cdots,-s$. In this process, the particle numbers $\hat
N_m$ with different $m$ are invariant. As a consequence, the total
spin $\hat{\boldsymbol S}$ and total particle number $\hat N$ are
also conserved. The other scattering process is $|s,m;s,-m\rangle
\langle s,m';s,-m'| + h.c.$ provided by the projector operator $\hat
P^{0}_{ab}$, as shown in Fig. \ref{fig1}(b). In this process, two
particles with opposite spins scatter into another pair, and the
absolute value $|m'|$ can be unequal to $|m|$. Obviously, this
process does not affect the total particle number $\hat N$, but it
destroys the invariability of $\hat N_m$. Therefore, the particle
number $\hat N_m$ no longer conserves. Nevertheless, by careful
consideration, we find that $\hat J_m = \hat N_m-\hat N_{-m}$ with
$m=s, s-1, \cdots$ and $m\geq 0$ are still invariant.  The invariance of $J_m$
means that the total spin $\hat {S} = \sum_m m\hat J_m$ is still a
good quantum number. Besides, some of the magnetic multipole
operators are also invariant. The multipole operators are observable
physical quantities and can be defined in the form of irreducible
tensors,
\begin{equation}
\label{mult}
\begin{split}
&\hat T^{l}_m \!=\!\sqrt{\!{{(l\!+\!m)! (l\!-\!m)!}}/\!{{(2l)!}}}
\Big(\!\prod_{j=1}^{l-1} \sum_{m_j}~\!^{^{\!\!'}} \! \hat {\tilde s}_{m_j}\! \Big)
\!\hat {\tilde s}_{m\!-\!\sum_{i=1}^{l-1} m_i},\\
&\hspace{53pt}m=-l,-l+1,\cdots,l,~~ l=1,2,\cdots,2s,\\
&\hat {\tilde s}_{-1} \!=\! (\hat s^x \!\!-\! {\rm i} \hat s^y)/\sqrt{2},
~ \hat {\tilde s}_0 \!=\!  {\sqrt{2}} \hat s^z
,~\hat {\tilde S}_1 \!=\! -(\hat s^x \!\!+\! {\rm i} \hat s^y)/\sqrt{2}.
\end{split}
\end{equation}
Here, $\hat s^{\alpha}$ ($\alpha=x,y,z$) is the spin operators of
one particle, $m_j=-1,0,1$,  and the sum $\Sigma'$ means
$|m-\sum_i^jm_i|<l-j$ for any $j$. The total multipole operators for
$N$ particles are $\hat M^l_m = \sum _{i=1}^N \hat T^{i,l}_m$, where
$\hat T^{i,l}_m$ is the $l$-rank multipole operator of the $i$-th
particle. It can be proved that the multipole operators with odd
rank are commutative with the  Hamiltonian
\begin{eqnarray}
\label{VM}
&& [\hat V, \hat M^l_m]=0,
\end{eqnarray}
where $\hat V=\sum_{i\neq j}^N (c_1\hat P^0_{ij} + c_2 \sum_{l=2,4, 
\cdots} \hat P^l_{ij})$, and thus are the conserved quantities of 
system (\ref{H}). This can be understood from the two-body scattering 
processes. If we only consider the process of permutation, all 
multipole operators are commutative with the spin part of the 
Hamiltonian, since the exchanges of spins have no effect on the 
magnetic  properties. However, the scattering processes of $\hat 
P^{0}_{ab}$ make the magnetic quadrupole change.

The odd rank multipole operators can be used to construct the 
generators of $Sp(2s+1)$ (half odd $s$) and $SO(2s+1)$ (integer $s$) 
algebras. Because the algebras $\mathfrak{so}(s,s+1)$ and
$\mathfrak{so}(2s+1)$ have the same complex extensions 
$\mathfrak{so}(2s+1, \mathds C)$, and if a model possesses the
$SO(s,s+1)$ symmetry, it must have the $SO(2s+1)$ symmetry. 
Here we use group $SO(s,s+1)$ instead of $SO(2s+1)$ to reveal 
the hidden symmetry in the system (\ref{H}). The generators 
of these groups can be defined uniformly as the $(2s+1) \times(2s+1)$
matrices satisfying the following conditions
\begin{eqnarray}
\label{gener}
&&\!\!\! {\rm tr}\hat Y=0,~ \hat Y=-\hat Y^*,~\hat Y^{t} \hat J+\hat J \hat Y=0.
\end{eqnarray}
Although the operator $\hat J$ is different for $Sp(2s+1)$ and
$SO(s,s+1)$ groups, we can write it in a uniform representation
\begin{eqnarray}
\label{J}
J^m_{m'} = (-1)^{s-m'}\delta_{m,-m'},~ m,m'=s,s-1,\cdots,-s.
\end{eqnarray}
Note that it is not valid for $SO(s,s+1)$
group with half odd  $s$, and we use $SO(s,s+1)$ group for the ones
with integer $s$ ($B_n$ type algebra, $n=s$) in the following
discussions.

Eq. (\ref{J}) shows a similarity of $Sp(2s+1)$ and $SO(s,s+1)$
groups. This enables the different symmetries of Hamiltonian
(\ref{H}) to be proven simultaneously. Since the multipole operators
defined in eq. (\ref{mult}) are all $(2s+1)^2-1$ linearly
independent $(2s+1)\times(2s+1)$ real matrices with zero trace, we
can use them to construct the generators of $Sp(2s+1)$ and
$SO(s,s+1)$ group. If we choose the basis of multipole operators
(\ref{mult}) as the eigenstates of $\hat s^z$ and the representation
of $\hat J$ takes the form of (\ref{J}), the generators of
$Sp(2s+1)$ and $SO(s, s+1)$ groups defined in (\ref{gener}) can be
represented as
\begin{eqnarray}
\hat Y^l_m = {\rm i} \hat T^j_m, \mbox{(odd $l$)}.
\end{eqnarray}
Therefore, $(s+1)(2s+1)$ operators $\hat Y^l_m$ are the generators
of $Sp(2s+1)$ group if $s$ is half odd, and the $s(2s+1)$ operators
$\hat Y^l_m$ are the generators of $SO(s,s+1)$ group if $s$ is
integer. For that multipole operators with odd rank-$l$ are
commutative with the spin part of Hamiltonian, eq. (\ref{VM}), the
corresponding symmetries hold for the system (\ref{H}). There is
also a $U(1)$ symmetry for the coordinate part, then the system
(\ref{H}) has $U(1)\otimes Sp(2s+1)$ symmetry for the fermionic case
and $U(1)\otimes SO(2s+1)$ symmetry for the bosonic case.

There are three homomorphisms $Sp(2)\simeq SU(2)$, $SO(3)\simeq
SU(2)$ and $Sp(4)\simeq SO(5)$. For the case $s=1/2$, only one
channel $\hat P^0_{ab}$ involved, the model in the spin sector has
the $SU(2)$ ($Sp(2)$) symmetry. For the cases $s=1$ and $3/2$, two
channels $\hat P^0_{ab}$ and $\hat P^2_{ab}$ are involved. When
$s=1$, the model has $SU(2)$ ($SO(3)$) symmetry, and when $s=3/2$
the system has $SO(5)$ ($Sp(4)$) symmetry, which are consistent with
the results obtained in the references \cite{WuCJ2003PRL}.

When $c_1=c_2$, the symmetry of the system (\ref{H}) in the spin
sector degenerates into the $SU(2s+1)$ one, where all the
interaction strengthes in different channels are the same. The
interaction of the spin part is the spin permutation operator up to
a constant. The permutation operator acting on the symmetric wave
functions gives the eigenvalue 1, and the one acting on the
anti-symmetric wave functions gives $-1$. Thus the effective
interaction is just the contact interaction and all magnetic
multipoles are conserved. This can also be explained from the view
that only the permutation operators are involved in the scattering
process. In the form of multipole operators, the $(2s+1)^2-1$
generators of $SU(2s+1)$ group read $\hat Y^s_{jm}\!=\! {\rm i}^m
(\hat T^j_m\!+\! \hat T^j_{\!-m})/2$ and $\hat Y^a_{jm}\!=\! {\rm
i}^{m+1} (\hat T^j_m\!-\!\hat T^j_{\!-m})\!/2$.

\section{Exactly solvable models}

In one dimension, it is well-known that at the $SU(2s+1)$ symmetry
point, the model is integrable. As we showed in the spin-1
\cite{Cao2007EPL} and spin-3/2 \cite{Jiang2009EPL} cases, there is
indeed another integrable point. For the $Sp(2s+1)$ or
$SO(2s+1)$-invariant Hamiltonian (\ref{H}), we construct the
following exactly solve model by constricting the parameters $c_0$
and $c_2$:
\begin{equation}
\label{Hi}
\begin{split}
&\hat H^{\rm int}=-\sum_{i=1}^N \frac{\partial^2}{\partial
{x}_i^2} + \sum_{i\neq j}^N \hat V^{\rm int}_{ij}
\delta({x}_i-{x}_j),\\
& \hat V^{\rm int}_{ij} =(-1)^{2s+1}\big[s +\frac{1}{2}-
(-1)^{2s}\big]c \hat P^0_{ij} + c \!\!\!\sum_{l=2,4, \cdots}\!\!\! \hat P^l_{ij}.
\end{split}
\end{equation}

With standard coordinate Bethe ansatz method, the wave function of
the system (\ref{Hi}) is assumed as
\begin{eqnarray}
\label{psi} 
\varPsi_{\!E} \!=\! \sum_{\mathcal{QP}} \!\Theta(x_{\!{\mathcal Q}_1}
\!<\! \cdots\!<\! x_{\!{\mathcal Q}_N}) A_{m_{1}\!, \cdots, m_{N}}
\!(\!\mathcal Q,\! \mathcal P){\rm e}^{{\rm i}\! \sum_{n}\! 
x_{\mathcal Q_n} k_{\mathcal P_n}}\!.\!\!\!\!
\end{eqnarray}
Here, $m_i$ is the spin component along $z$-direction of $i$-th
particle, $m_i=s,s-1,\cdots,-s$, and $k_i$ ($i=1, 2, \cdots, N$) are
the quasi-momenta carried by the particles. ${\mathcal Q}$ and
${\mathcal P}$ are all $N!$ permutations of $\{1,2,\cdots, N\}$, and
${\mathcal Q}_i$ ( ${\mathcal P}_i$) is the $i$-th number of the
permutation ${\mathcal Q}$ (${\mathcal P}$). $\Theta(x_{{\mathcal
Q}_1}\!<\!\cdots\!<\!x_{{\mathcal Q}_N})\!=\!\prod_{i=2}^N
\theta(x_{{\mathcal Q}_i}\! -\! x_{{\mathcal Q}_{i-1}})$ is
continuous multiplication of step function $\theta(x)$. When
$x\geq0$, $\theta(x) = 1$, and otherwise $\theta(x) = 0$. Thus the
function $\Theta$ divides the coordinate space into $N!$ intervals.

The two-particle scattering occurs at the interface of two adjacent
coordinate intervals ${\cal Q}$ and ${\cal Q'}$. The scattering
matrix of particles $a$ and $b$ carrying different quasi-momenta is
defined in the two-particle spin space to describe the relation of
the superposition coefficients
\begin{eqnarray}
&& \hspace{-20pt}\vec A({\mathcal
Q},{\mathcal P})= \hat S_{^{{\mathcal
Q}_\xi,{\mathcal Q}_{\xi+1}}}(k_{^{{\mathcal
P}_{\xi+1}}}\!\!-\!k_{^{{\mathcal P}_\xi}})
\vec A({\mathcal Q}',{\mathcal P}'),\\
&& \hspace{-20pt} {\mathcal Q}'=\{{\mathcal Q}_1,{\mathcal Q}_2,\cdots,
{\mathcal Q}_{\xi-1}, {\mathcal Q}_{\xi+1}, {\mathcal Q}_{\xi},
{\mathcal Q}_{\xi +2} \cdots,{\mathcal Q}_N\},\!\! \nonumber\\
&& \hspace{-20pt} {\mathcal P}'=\{{\mathcal P}_{~\!1},{\mathcal P}_{~\!\!2} ,\cdots,
{\mathcal P}_{~\!\xi-1}, {\mathcal P}_{~\!\!\xi+1}, {\mathcal P}_{~\!\xi},
{\mathcal P}_{~\!\!\xi +2} \cdots,{\mathcal P}_N\}, \nonumber
\end{eqnarray}
where $\xi=1,2, \cdots, N-1$, ${\cal Q}_\xi=a$, ${\cal Q}_{\xi+1}=b$
and $\vec A$ is the vector denotation of superposition coefficients
$A_{m_1,\cdots,m_N}$. In the system (\ref{Hi}), the wave function
should be continuous and the first-order derivative of the wave
function with respect to coordinate should be discontinuous. Solving
the Sch\"odinger equation and using the symmetry or antisymmetry
condition, we can obtain the scatting matrix. For the
$Sp(2s+1)$-invariant fermionic model, the two-body scattering matrix
is
\begin{eqnarray}
\label{spn_S} \hat S^{(s)}_{ab} \!(\lambda)\! = \!\!\!
\sum_{l=1,3,\cdots} ^{2s}\!\!\! \hat P^l_{ab} \!+\!\!\!
\sum_{l=2,4,\cdots} ^{2s-1}\!\!\! \frac{\lambda \!-\!{\rm
i}c}{\lambda \!+\! {\rm i}c} \hat P^l_{ab} \!+\! \frac{\lambda \!-\!
(s \!+\!\frac32) {\rm i}c}{\lambda \!+\!(s\!+\!\frac32){\rm i}c}
\hat P^0_{ab}.\!\!
\end{eqnarray}
For the $SO(2s+1)$-invariant bosonic model, the scattering matrix
reads
\begin{eqnarray}
\label{son_S}
\hat S^{(s)}_{ab}\!(\lambda)  \!=\!\!\!
\sum_{l=1,3,\cdots} ^{2s-1}\!\!\! \hat P^l_{ab} +\!\!\! \sum_{l=2,4,\cdots} ^{2s} \!\!\!
\frac{\lambda\!-\!{\rm i}c} {\lambda\!+\!{\rm i}c} \hat P^l_{ab} \!+\!
\frac{\lambda\!+\!(s\!+\!\frac12){\rm i}c}
{\lambda\!-\! (s\!+\!\frac{1}{2}) {\rm i}c} \hat P^0_{ab}.\!\!
\end{eqnarray}

The scattering matrices (\ref{spn_S}) and (\ref{son_S}) are
different. In order to prove the integrability of the bosonic and
fermionic models uniformly, we introduce the $R$-matrix for these
two kinds of symmetries by the following mapping,
\begin{equation}
\label{Rs} R^{(s)}_{ab} \!(\lambda) \!=\!\! \left\{\!\!\!
\begin{array}{ll}-\hat P_{ab} S^{(s)}_{ab} (\lambda), & (\mbox{half
odd $s$}), \\{} [a(\lambda)\!-\! b(\lambda)] \hat P_{\!ab}
S^{(s)}_{ab} (-\lambda),\!\! &(\mbox{integer $s$}).
\end{array} \right.
\end{equation}
With this mapping, the explicit form of $R$-matrix is
\begin{eqnarray}
\label{Rthis} \hat R^{(s)}_{ab}(\lambda) =
b(\lambda)\hat I + a(\lambda)\hat P_{ab} -
(2s+1)z^{(s)}(\lambda)\hat P^0_{ab}.
\end{eqnarray}
Here, $b(\lambda) = {\rm i}c/(\lambda+ {\rm i}c)$, $a(\lambda) =
\lambda/ (\lambda+{\rm i}c)$, $\hat I$ is a unitary operator and
$z^{(s)}(\lambda)$ is a scalar function depending on $s$,
\begin{eqnarray}
\label{zs} z^{(s)}(\lambda) = (-1)^{2s} b(\lambda) a(\lambda/
[s+1/2 +(-1)^{2s+1}]).
\end{eqnarray}
After some calculations, we find that for any spin-$s$, integer or
half odd, $\hat R^{(s)}$ satisfies the Yang--Baxter equation
\begin{eqnarray}
\label{ybe} \hat R_{ab}(\lambda)\hat R_{bc}(\lambda+\mu)\hat
R_{ab}(\mu) =\hat R_{bc}(\mu)\hat R_{ab}(\lambda+\mu)\hat
R_{bc}(\lambda).
\end{eqnarray}
In the derivation, the following relations have been used
\begin{eqnarray}
\hat P^0_{ab} \hat P^{~}_{bc}=(2s+1)\hat P^0_{ab}\hat P^0_{ac}, ~ \hat P_{ab} =
\sum_{l=0}^{2s}(-1)^{2s-l} \hat P^l_{ab}.
\end{eqnarray}
Since there are two invariant mappings, $R(\lambda)\mapsto
f(\lambda)R(\lambda)$ and $R(\lambda)\mapsto R(-\lambda)$, for the
Yang--Baxter equation (\ref{ybe}) \cite{Kulish1982JMS}, Hamiltonian
(\ref{Hi}) is integrable.

$R$-matrix defined in eq. (\ref{Rthis}) only has two sets of
solutions of the Yang--Baxter equation (\ref{ybe}). One set is
$z^{(s)}=0$ where the system has the $SU(2s+1)$ symmetry. In this
case, there are no effective spin exchange interactions. The other
set is eq. (\ref{zs}). The system has the $Sp(2s+1)$ symmetry for
half odd $s$ and the $SO(2s+1)$ symmetry for integer $s$. The
corresponding integrable spin chains are Kennedy--Batchelor models
in \cite{Kennedy1992JPA}. When $s=1/2$, the $Sp(2)$-invariant
integrable model is discussed in \cite{Yang1967PRL};
when $s=1$, the $SO(3)$-invariant integrable model is discussed in
\cite{Cao2007EPL}; and when $s=3/2$, the $Sp(4)$-symmetry integrable
model is discussed in \cite{Jiang2009EPL}.

The integrable model (\ref{Hi}) has one tunable interacting
parameter $c$. For the $Sp(2s+1)$ fermionic model, the interaction
is repulsive when $c>0$ and is attractive when $c<0$. For the
$SO(2s+1)$ bosonic model, the interaction in spin-0 channel is
attractive and that in other channels is repulsive when $c>0$, while
the interaction in the spin-0 channel is repulsive and is attractive
in other channels when $c<0$. To obtain the exact energy spectrum of
the system, we need to determine all the values of quasi-momenta
$k_j$. This can be done by solving the eigenvalue problem given by
the periodic boundary condition, in which we can obtain the Bethe
ansatz equations.

\section{Exact solutions}

For the integrable systems with high symmetry, the exact solutions
are usually obtained by using the nested algebraic Bethe ansatz
method. The Bethe ansatz equations of integrable quantum gas models
are composed of the ones of the coordinate part, i. e. $U(1)$
symmetry and the ones given by the spin part. The spin sector
usually has nesting integrable  symmetries for high spin models.
Using the method suggested in \cite{Martins1997NPB, Martins2002NPB},
we can obtain the Bethe ansatz equations for the $Sp(2s+1)$ model
($C_n$ type algebra, $n=s+1$) and the $SO(2s+1)$ model.

For $Sp(2s+1)$ ($s>1/2$) case, there are $s+3/2$ sets of coupled
equations. When $s>3/2$, the equations are
\begin{eqnarray}
\label{Sp:BAEs-1} &&\hspace{-25pt} {\rm e}^{ik_jL} =
\prod_{i=1}^{M^{_{(s)}}} \frac{k_j-\lambda^{_{(s)}}_i+{\rm i}\frac
c2} {k_j-\lambda^{_{(s)}}_i - {\rm i}\frac c2}, \hspace{41pt}j =
1,2,\cdots, N, \\
\label{Sp:BAEs-2} &&\hspace{-25pt} \prod_{i=1}^
{M^{_{\!(l+1)}}\!\!}\!\! \frac{\lambda^{_{\!(l)}}_j \!\!-\!
\lambda^{_{\!(l+1)}}_i \!\!+\! {\rm i}\frac c2}
{\lambda^{_{\!(l)}}_j \!\!-\! \lambda^{_{\!(l+1)}}_i \!\!-\! {\rm
i}\frac c2} \! \prod_{i=1}^ {M^{_{\!(l-1)}\!\!}}\!\!
\frac{\lambda^{_{\!(l)}}_j \!\!-\! \lambda^{_{\!(l-1)}}_i \!\!+\!
{\rm i}\frac c2} {\lambda^{_{\!(l)}}_j \!\!-\!
\lambda^{_{\!(l-1)}}_i \!\!-\! {\rm i}\frac c2} \!=\!\!\!
\prod_{j'\neq j} ^{M^{_{\!(l)}}} \!\! \frac{\lambda^{_{\!(l)}}_j
\!\!-\! \lambda^{_{\!(l)}}_{j'} \!\!+\! {\rm i}c}
{\lambda^{_{\!(l)}}_{j} \!\!-\! \lambda^{_{\!(l)}}_{j'}\!\!
-\!{\rm i}c},\nonumber\\
&&\hspace{28pt}l = s,s-1,\cdots, 5/2,~~ j = 1,2, \cdots, M^{_{(l)}},\\
\label{Sp:BAEs-3}
&&\hspace{-25pt}
\prod_{i=1}^ {M^{_{\!(\!\frac52\!)}}}
\frac{\lambda^{_{\!(\!\frac32\!)}}_j \!-
\lambda^{_{\!(\!\frac52\!)}}_i + {\rm
i}\frac c2} {\lambda^{_{\!(\!\frac32\!)}}_j \!-
\lambda^{_{\!(\!\frac52\!)}}_i - {\rm
i}\frac c2}  \prod_{i=1}^ {M^{_{\!(\!\frac12\!)}}}
\frac{\lambda^{_{\!(\!\frac32\!)}}_j \!-
\lambda^{_{\!(\!\frac12\!)}}_i + {\rm
i} c} {\lambda^{_{\!(\!\frac32\!)}}_j \!- \lambda^{_{\!(\!\frac12\!)}}_i - {\rm
i}c} \!=\!\!\!
\prod_{j'\neq j} ^{M^{_{\!(\!\frac32\!)}}} \!\! \frac{\lambda^{_{\!(\!\frac32\!)}}_j
\!\!-\! \lambda^{_{\!(\!\frac32\!)}}_{j'} \!\!+\! {\rm i}c}
{\lambda^{_{\!(\!\frac32\!)}}_{j} \!\!-\! \lambda^{_{\!(\!\frac32\!)}}_{j'}\!\!
-\!{\rm i}c},\nonumber\\
&&\hspace{125pt} j = 1,2, \cdots, M^{_{(l)}},\\
\label{Sp:BAEs-4}
&&\hspace{-25pt}
\prod_{i=1}^ {M^{_{\!(\!\frac12\!)}}}\!\!
\frac{\lambda^{_{\!(\!\frac32\!)}}_j \!\!-\!
\lambda^{_{\!(\!\frac32\!)}}_i \!\!+\! {\rm
i}c} {\lambda^{_{\!(\!\frac12\!)}}_j \!\!-\!
\lambda^{_{\!(\!\frac32\!)}}_i \!\!-\! {\rm
i}c}  \!=\!\!\!
\prod_{j'\neq j} ^{M^{_{\!(\!\frac32\!)}}} \!\! \frac{\lambda^{_{\!(\!\frac32\!)}}_j
\!\!-\! \lambda^{_{\!(\!\frac32\!)}}_{j'} \!\!+\! 2{\rm i}c}
{\lambda^{_{\!(\!\frac32\!)}}_{j} \!\!-\! \lambda^{_{\!(\!\frac32\!)}}_{j'}\!\!
-\!2{\rm i}c},~j = 1,2, \cdots, M^{_{(l)}}.
\end{eqnarray}
Here, $M^{(l)}$ is the numbers of rapidity $\lambda^{(l)}$,
$M^{(2s+1)} \!=\! N$, and $\lambda^{(s+1)}_j \!=\! k_j$. When
$s\!=\!3/2$, the Bethe ansatz equations degenerate into the ones
obtained in \cite{Jiang2009EPL}. When $s=1/2$, the system (\ref{Hi})
degenerates into the $Sp(2)$-invariant spin-1/2 Fermi gas, and the
Bethe ansatz equations are given in \cite{Yang1967PRL}.

For the $SO(2s+1)$ bosons,  the Bethe ansatz equations have $s+1$
sets, and when $s>1$ they are
\begin{eqnarray}
\label{SO:BAEs-1}
&&\hspace{-25pt}
{\rm e}^{ik_jL} \!= \! \prod_{i\neq j}^N\!
\frac{k_j\!-\!k_i\!+\!{\rm i}c} {k_j\!-\!k_i\!
- \!{\rm i}c} \!\!
\prod_{i=1}^{M^{_{(s)}}}
\frac{k_j\!-\!\lambda^{_{(s)}}_i\!-\!{\rm i}\frac c2}
{k_j\!-\!\lambda^{_{(s)}}_i \!+\!{\rm i}\frac c2}, ~j \!=\!
1,2,\cdots\!, N,
\\
\label{SO:BAEs-2}
&&\hspace{-25pt}
\prod_{i=1}^ {M^{_{\!(l+1)}}\!\!}\!\!
\frac{\lambda^{_{\!(l)}}_j \!\!-\!
\lambda^{_{\!(l+1)}}_i \!\!-\! {\rm
i}\frac c2} {\lambda^{_{\!(l)}}_j \!\!-\!
\lambda^{_{\!(l+1)}}_i \!\!+\! {\rm
i}\frac c2} \! \prod_{i=1}^ {M^{_{\!(l-1)}\!\!}}\!\!
\frac{\lambda^{_{\!(l)}}_j \!\!-\!
\lambda^{_{\!(l-1)}}_i \!\!-\! {\rm
i}\frac c2} {\lambda^{_{\!(l)}}_j \!\!-\!
\lambda^{_{\!(l-1)}}_i \!\!+\! {\rm
i}\frac c2} \!=\!\!\!
\prod_{j'\neq j} ^{M^{_{\!(l)}}} \!\! \frac{\lambda^{_{\!(l)}}_j
\!\!-\! \lambda^{_{\!(l)}}_{j'} \!\!-\! {\rm i}c}
{\lambda^{_{\!(l)}}_{j} \!\!-\! \lambda^{_{\!(l)}}_{j'}\!\!
+\!{\rm i}c},\nonumber\\
&&\hspace{39pt}l = s,s-1,\cdots, 2,~~ j = 1,2, \cdots, M^{_{(l)}},\\
\label{SO:BAEs-3}
&&\hspace{-25pt}
\prod_{i=1}^ {M^{_{\!(2)}}}\!\!
\frac{\lambda^{_{\!(1)}}_j \!\!-\!
\lambda^{_{\!(2)}}_i \!\!-\! {\rm
i}c} {\lambda^{_{\!(1)}}_j \!\!-\!
\lambda^{_{\!(2)}}_i \!\!+\! {\rm
i}c}  \!=\!\!\!
\prod_{j'\neq j} ^{M^{_{\!(1)}}} \!\! \frac{\lambda^{_{\!(1\!)}}_j
\!\!-\! \lambda^{_{\!(1)}}_{j'} \!\!-\! {\rm i}c}
{\lambda^{_{\!(1)}}_{j} \!\!-\! \lambda^{_{\!(1)}}_{j'}\!\!
+\!{\rm i}c},~j = 1,2, \cdots, M^{_{(l)}}.
\end{eqnarray}
When $s=1$, the above Bethe ansatz equations degenerate into ones
obtained in \cite{Cao2007EPL}.

Therefore, if the quasi-momenta $k$'s satisfy the Bethe ansatz
equations, $\varPsi_E$ (\ref{psi}) is the eigen-wave-function of the
system and the corresponding eigenvalues of energy and momentum are
\begin{equation}
E = \sum_{j=1}^N k^2_j, \quad \quad K = \sum_{j=1}^N k_j.
\end{equation}
The total spin is $S = sN-\sum_{l} M^{(l)}$.

Obviously, the Bethe ansatz equations of the present system are
different from the $SU(2s+1)$ ones. The physical properties can be
obtained from the solutions of Bethe ansatz equations. For example,
solutions of $SO(3)$-invariant spin-1 bosonic model show that there
are bound states in the regimes of $c>0$ and $c<0$
\cite{Cao2007EPL}, for that there always exist attractive
interactions in some scattering channels.

\section{Repulsive fermions}

For the repulsive fermionic models, detailed analysis of the Bethe
ansatz equations shows that all quasi-momenta $k$ are real, which
means there are no charge bound states, and the spin rapidities
$\lambda^{(l)}$ form strings. In the thermodynamic limit, the string
solutions read \cite{Martins2002NPB}
\begin{eqnarray}
&& \hspace{-20pt} \lambda^{(l)}_{n,z,j} =  \lambda^{(l)}_{n,z}+(n+1-2j){\rm i}c/2,
~ j=1,2,\cdots,n,\nonumber\\
&&\hspace{138pt}3/2\leq l \leq s,\\
&&\hspace{-20pt} \lambda^{(1\!/\!2)}_{n,z,j} =
\lambda^{(1\!/\!2)}_{n,z}+(n+1-2j){\rm i}c,~~~\!j=1,2,\cdots,n.
\end{eqnarray}
Here, $\lambda^{(l)}_{n,z}$ denote the real parts of the $n$-string
rapidities, $z=1,2,\cdots , M^{(k)}_n$, and $M^{(k)}_n$ is the
number of $n$-strings for $\lambda^{(l)}$. Based on the above string
hypothesis, the finite temperature thermodynamic properties of the
system can be obtained. If the temperature tends to zero, only the
real rapidities and 2-strings for $\lambda^{(l)}$ ($3/2\leq l\leq
s$) are left in the ground state. Substituting these solutions into
the Bethe ansatz equations and taking the thermodynamic limit, we
obtain the coupled integral equations. Solving these equations, we
obtain the numbers of the $i$-string $\lambda^{(l)}$ analytically
\begin{equation}
M^{(l)}_1 \!\!=\!\! \frac{l\!-\!\frac12}{s\!+\!\frac12}N,
M^{(l)}_2\!\!=\!\! \frac{s\!-\!l\!+\!1}{2(s\!+\!\frac12)}N,
(l\!>\!\frac{_3}{^2}),~\!M^{(\!\frac12\!)}_1\!\!=\!\!\frac12N.
\end{equation}
Thus the numbers of $\lambda^{(l)}$ are $M^{(l)}=N$($l>3/2$),
$M^{(\frac12)}=N/2$, and the conserved quantities $J_m=0$ in the
ground state. The total spin is zero, so that the ground state is
spin singlet state. Since the string distributions are symmetric
around the real axis, the total momentum $K$ of the grounds state is
zero.

\begin{figure}[t]
\begin{center}
\includegraphics[width=\linewidth]{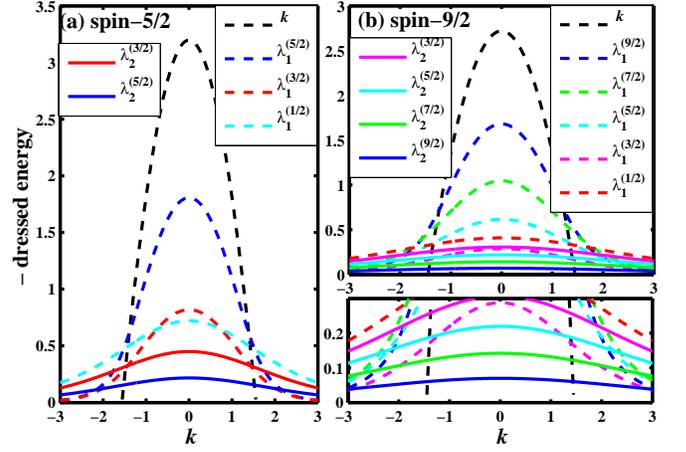}
\end{center}
\caption{The ground state dressed energies of spin-5/2 (a) and spin-9/2
(b). Here $c=1$ and $n=1$. \label{fig2}}
\end{figure}

The dressed energy of charge rapidities $k$ in the ground state
satisfies the following equations,
\begin{eqnarray}
\label{dress} &&\epsilon(k) = k^2- \mu + \hat{ D}^{(s+1)}
*\epsilon(k), ~~|k|< Q, \nonumber \\
&&\epsilon^{(l)}_{i} (k)= \hat{D}^{(l)}_i *\epsilon(k), \quad
l=\frac{1}{2}, \cdots, s.
\end{eqnarray}
Here $\mu$ is the chemical potential, $*$ is an integral operation
defined by $ \hat w*y(x)=\int w(x-x') y(x) {\rm d} x'$, $Q$ is the
Fermi point which is determined by the particle density $n=N/L$, and
the kernels $D(k)$ of integral operators $\hat{{D}}$ are
\begin{equation}
\begin{split}
&{D} (k) = \hat {{a}}_{1/2} * {D}^{(s)}_1(k) +\hat {a}_{1}* {{D}}^{(s)}_2(k),\\ 
&{{D}} ^{(l)}_1(k) = \frac{1}{2s+1} \frac{\sin(\frac{2l-1}{2s+1}\pi)} {\cosh(\frac{2\pi k/c}{2s+1}) +\cos(\frac{2l-1}{2s+1}\pi)},\\
&{{D}} ^{(l)}_2(k) = \hat{{D}}^{(s-l+3/2)}_1 *{D}_1^{(1/2)}
(k),~ {3}/2\leq l \leq s,\\
&{{D}} ^{(1/2)} (k) ={1}/[(2s+3)\cosh[{\pi k}/{(c(s+3/2))}]],
\end{split}
\end{equation}
where $a(x)=t/[\pi(x^2+(tc)^2)]$. The dressed energy for $c=1$ and
$n=1$ is shown in Fig. \ref{fig2}.

The physical properties of such 1D systems are controlled by the
parameter $\gamma = c/n$ \cite{LiebLiniger1963PR}. When $\gamma\to
\infty$, we obtain the density of states, energy and Fermi point in
the strong repulsive limit as
\begin{equation}
\rho(k)=\frac{1}{2\pi},(k\!\leq\!|Q|),~~ E=\frac{1}{3\pi}Q^3,~~Q= n \pi.
\end{equation}
When $\gamma \to 0$, the system degenerates into the free fermions
and we have
\begin{equation}
\rho(k)\!=\!\frac{2s\!+\!1}{2\pi},(k\!\leq\!|Q|),~ E\!=\!\frac{2s\!+\!1}
{3\pi}Q^3,~Q\!=\! \frac{n\pi}{2s\!+\!1}.\!
\end{equation}

\section{Conclusion}
In conclusion, we find that there is a hidden symmetry of the high
spin cold atomic systems with a special interaction form away from
the $SU(2s+1)$ symmetry point. Based on the symmetry analysis, a
new class of integrable models for cold atoms with arbitrary spin is
proposed.

\section{Acknowledgments} We would like to thank Prof. Shu Chen, 
Xi-Wen Guan, Zhong-Qi Ma, M. T. Batchelor and G. V. Shlyapnikov for
the beneficial discussions. This work was supported by the NSFC, 
the Knowledge Innovation Project of CAS, and the National Program
for Basic Research of MOST.

* Email: yupeng@iphy.ac.cn

\end{document}